\documentclass[aps,pre,twocolumn,groupedaddress,10pt,showpacs,showkeys,superscriptaddress]{revtex4-1}
\usepackage{amsmath,amssymb}
\usepackage{graphicx,color}

\newcommand{\dif}{\mathrm{d}}%
\newcommand{\Eins}{\mathbf{1}}%
\newcommand{\fdif}{\operatorname{\delta}}
\newcommand{\Fdif}[2]{\frac{\fdif\!#1}{\fdif\!#2}}
\newcommand{\ii}{i}%
\newcommand{\Nabla}{\vec{\nabla}}%
\newcommand{\pdif}[2]{\frac{\partial#1}{\partial#2}}%
\newcommand{\R}{\mathbb{R}}%
\newcommand{\diag}{\operatorname{diag}}%

\begin{document}
\title{Dynamical density functional theory for colloidal particles with arbitrary shape}

\author{Raphael Wittkowski}\author{Hartmut L{\"o}wen}
\thanks{Invited contribution to the Special Issue of Molecular Physics in honor of Luciano Reatto}
\affiliation{Institut f{\"u}r Theoretische Physik II, Weiche Materie,
Heinrich-Heine-Universit{\"a}t D{\"u}sseldorf, D-40225 D{\"u}sseldorf, Germany}
\date{\today}

\begin{abstract}
Starting from the many-particle Smoluchowski equation, we derive dynamical density functional theory for 
Brownian particles with an arbitrary shape. 
Both passive and active (self-propelled) particles are considered.
The resulting theory constitutes a microscopic framework to explore the collective dynamical behavior of biaxial particles 
in nonequilibrium. For spherical and uniaxial particles, earlier derived dynamical density functional theories 
are recovered as special cases. 
Our study is motivated by recent experimental progress in preparing colloidal particles with many 
different biaxial shapes.
\end{abstract}


% Colloids: 82.70.Dd
% Brownian motion: 05.40.Jc
% Theory and models of liquid crystal structure: 61.20.Gy 
% Molecular and microscopic models and theories of liquid crystal structure: 61.30.Cz 

\maketitle

%**************************************************************************
%**************************************************************************

\section{\label{sec:introduction}Introduction}
In its original form, classical dynamical density functional theory (DDFT) was derived by Marconi and Tarazona \cite{MarconiT1999}
in 1999 for spherical, i.e., isotropic, colloidal particles.
Their derivation started from the Langevin equation for spherical particles \cite{Dean1996} that interact via a pair potential. 
Later, in 2004, DDFT was rederived by Archer and Evans \cite{ArcherE2004} from the Smoluchowski equation that corresponds to the 
Langevin equation for interacting spherical particles. 
In 2007, DDFT was generalized by Rex, Wensink, and L\"owen \cite{RexWL2007} to systems of uniaxial anisotropic particles with 
orientational degrees of freedom.
This generalization is based on the Smoluchowski equation for rigid rods \cite{Dhont1996}.
It made DDFT applicable to the important class of uniaxial liquid crystals. 
%
% However, there are also biaxial liquid crystals like the banana-shaped bent-core mesogens \cite{Vorlaender1929,VorlaenderA1932}, 
% to which this DDFT is not applicable.  
% Due to their unique electro-optical properties including very fast ferroelectric switching, this class of liquid crystals is very important 
% for liquid crystal display technology and gains growing interest in science and technology. 
% Especially the observation that banana-shaped liquid crystalline particles exhibit ferroelectric phases \cite{NioriSWFT1996} 
% and the discovery of a biaxial nematic liquid crystalline phase \cite{AcharyaPK2004,MadsenDNS2004} 
% led to a big increase of scientific work in this area.

Nowadays, it is already possible to produce colloidal particles with rather complicated shapes including biaxial particles. 
%Beside the traditional spheres, colloidal rods, and, for example, dumbbells, even nanocubes \cite{ZhouLXWZZW2006}, nanostars \cite{WuCH2009}, 
%multipod-shaped nanocrystals \cite{NewtonW2007,DekaMDGBM2010}, and star-shaped dendrites \cite{ZhouLXWZZW2006} can be synthesized now.
Although static classical density functional theory (DFT) has presently available very powerful tools like fundamental measure theory \cite{HansenGoosM2009}
that allow to consider also such complicated colloidal particles in the context of DFT, the dynamics of these biaxial particles
could up to now not be investigated on the basis of DDFT.  
For these reasons, it is of high importance to push forward the development of DDFT.

In this paper, we present a further generalization of DDFT, which is now also applicable to biaxial particles. 
This extension of DDFT contains the previous DDFT equations as special cases and does not assume a certain shape for the colloidal particles. 
Instead, it is derived for arbitrarily shaped colloids. In comparison with the former DDFT approach, this leads to three independent 
rotational diffusion coefficients instead of only one. 
Since our new DDFT equation holds also for screw-like particles, it takes even a possible translational-rotational coupling into account.
Additionally, we consider a possible self-propulsion mechanism of the particles so that our results are also relevant for the investigation 
of the collective dynamics of active particles like swarms of swimming microorganisms as, for example, protozoa \cite{Ramaswamy2010}. 

The paper is organized as follows: after giving a short overview in Sec.\ \ref{sec:classification} about anisotropic colloidal particle shapes 
that can already be synthesized, we present our derivation of the extended DDFT equation in Sec.\ \ref{sec:derivation} and discuss 
special cases that are known from literature. Sec.\ \ref{sec:applications} is addressed to possible applications of the DDFT equation.  
Finally, we give conclusions and mention possible further extensions of DDFT in Sec.\ \ref{sec:conclusions}.

\section{\label{sec:classification}Geometric classification of colloidal particles}
\begin{figure*}[ht]
\centering
\includegraphics[height=\linewidth]{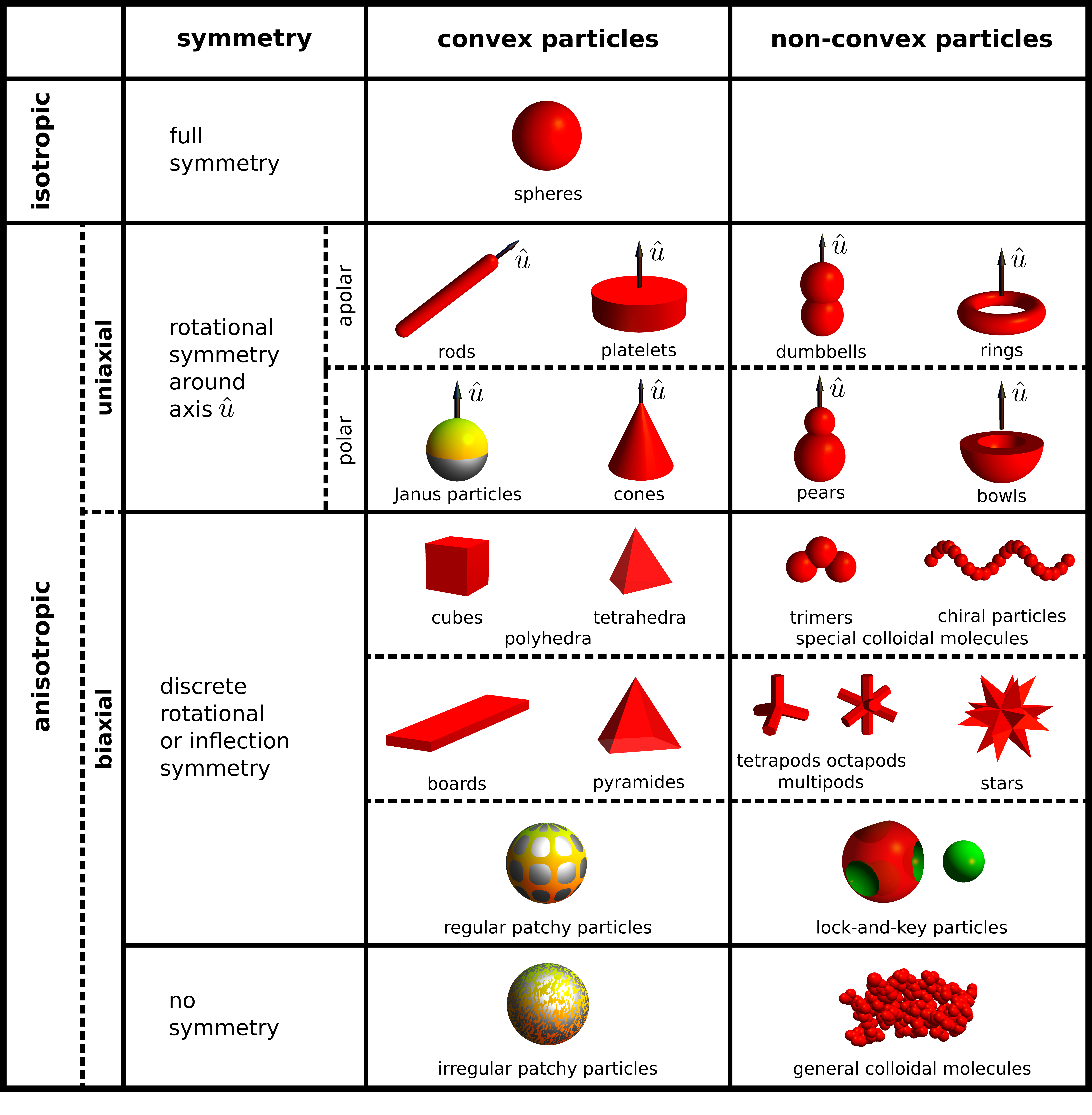}
\caption{Classification of synthetizable colloidal particles with respect to their shape. 
Geometrical properties that were used to classify the shapes are symmetry and convexity.}
\label{fig:cc}
\end{figure*}

Induced by technological advance in the processing of nanomaterials, a large number of differently shaped colloidal particles became 
synthetizable during the last years. 
The different shapes of these colloidal particles can be classified by means of their geometric properties.
Figure \ref{fig:cc} shows a detailed classification of colloidal shapes with respect to symmetry and convexity.
Such a classification is of big importance since colloidal particles may form a huge set of mesotropic phases (mesophases) 
\cite{Chandrasekhar1992,Boyd2008K56} that go along with different states of translational and orientational order. 
The possible states of translational and orientational order depend strongly on the shapes of the particles and a classification of their 
shapes is therefore also a classification of the possible phases that these particles may evolve.

The most simple and at once full symmetric, i.e., \emph{isotropic}, shape is the sphere. 
This is the traditional shape for colloids in theoretical soft matter physics, because it is simple to produce and due to a lack of orientational
degrees of freedom relatively simple to describe theoretically. 
Since spheres possess only translational degrees of freedom, they solely appear in the completely disordered isotropic phase and
in the crystalline state \cite{HansenMD2006}. The shape of a sphere is globally convex and there is no non-convex analog with full symmetry.
All other colloidal particle shapes are \emph{anisotropic} and either uniaxial or biaxial.
The characteristic property of \emph{uniaxial} particles is a symmetry axis, whose orientation is denoted by the unit vector $\hat{u}$ 
in the following. These particles have rotational symmetry and possess one orientational degree of freedom in two spatial dimensions and 
two orientational degrees of freedom in three spatial dimensions. 
Uniaxial particles are further distinguished into apolar and polar particles. 
An uniaxial particle is called \emph{apolar}, if it has head-tail symmetry and \emph{polar} otherwise.
Rod-like particles \cite{RoordavDPGvBK2004,HanANZLY2006} like spherocylinders, spheroids, and ellipsoids are the most simple anisotropic colloidal particles. 
They are convex and apolar and of big importance since they may evolve the industrially important nematic phase and serve as excellent
model systems for most liquid crystals \cite{WeyerichDACK1990,TavaresHTdG2009,LopezLRPC2010}. 
A further member of convex and apolar particles are the platelets \cite{WierengaLP1998,BierHD2005,vanderBeekEtAl2006,vanderBeekDWVL2008,LapointeHMS2010}. 
They have a similar phase diagram to rod-like particles with a strong affinity to form columnar stacks \cite{MouradPVL2010}. 
Systems of such disk-like particles are realized in nature for example by clay suspensions \cite{DijkstraHM1995,MourchidDLLL1995,DijkstraHM1997}. 
Examples for non-convex apolar particles are dumbbells (dimers) \cite{JohnsonvKvB2005,MarechalD2008,DemiroersJvKvBI2010}, 
that are produced by mergence of two spheres of equal size, and rings \cite{YanLWLZ2006,Thaokar2008}, 
that can be made by etching from colloidal spheres that are partially embedded in a metal layer. 
The complement of apolar particles is built by the \emph{polar} particles, that have no head-tail symmetry.
A famous member of this particle class are the Janus particles \cite{HongCLG2006,HoCSLK2008,WaltherM2008}.
%, who were named by Christiane Casagrande et\,al.\ in 1988 after the Roman god Janus \cite{CasagrandeV1988,CasagrandeFRV1989}. 
They are spheres with a different coating at one half of the surface. The original Janus particles had a hydrophilic and a hydrophobic coating. 
Nowadays, one coating is often reactive like a platinic coating that decomposes hydrogen peroxide catalytically. 
Such particles are immersed into a hydrogen peroxide solution to realize active particles (micro-swimmers) that are driven by an 
intrinsic drive \cite{FattahLLGWLBA2011}. Cones are a further member of uniaxial polar particles.  
Carbon nanocones appear naturally in graphite \cite{KrishnanDTHLE1997,JaszczakRDG2003,NalumNaessEHK2009} and do not need to be produced by an 
elaborate method. 
By the mergence of two spheres with different diameters, one obtains a pear-like particle \cite{KegelBEP2006,HoseinJLEL2009}. 
Pears and also bowls \cite{JeongICKX2007,MarechalKDID2010} are non-convex particles that are uniaxial and polar. 
The latter stack into each other and form columnar structures \cite{MarechalD2010}.  

Particles with less symmetry are \emph{biaxial}. They are the complement to the uniaxial particles in the class of the anisotropic particles. 
Biaxial particles have either only discrete symmetries, like inflection symmetry and discrete rotational symmetry, 
or are completely asymmetric. 
In both cases, the biaxial particles have three orientational degrees of freedom and a unit vector is no longer sufficient to describe 
their orientation.
Instead, two perpendicular unit vectors or Eulerian angles have to be used \cite{DoiE2007}.
Due to the additional orientational degree of freedom, the phase diagrams of biaxial colloidal particles are much richer than those for 
uniaxial particles \cite{SircarLW2010}.
Convex colloidal particles with discrete rotational or inflection symmetry are, for example, polyhedra like cubes 
\cite{CuestaMR1997,MartinezRatonC1999,ZhouLXWZZW2006,SevonkaevGM2008} and 
tetrahedra \cite{YinW1997,HajiAkbariEKZPPMG2009}, boards \cite{vandenPolPTWBV2009}, pyramids \cite{GuhaKC2004,Helseth2005,FanTOB2008}, 
and regular patchy particles \cite{ZhangG2004,BianchiLTZS2006,ChoYKJEYYP2007,Sciortino2008}. 
The latter differ from Janus particles by a patchy coating with a regular, e.\ g., tetrahedral, arrangement.
Non-convex particles with discrete rotational or inflection symmetry include special colloidal molecules that are realized by more than two
spheres that are merged in a regular arrangement. Examples for this include trimers \cite{LiddellS2003} consisting of three equal spheres and 
chiral particles \cite{ZerroukiBPCB2008,WensinkJ2009} consisting of many equal spheres in a helical arrangement, 
multipod-shaped nanocrystals \cite{NewtonW2007,DekaMDGBM2010}, stars \cite{ZhouLXWZZW2006,WuCH2009}, 
and some lock-and-key particles \cite{SacannaICP2010}.
% Also the banana-shaped bent-core mesogens \cite{Vorlaender1929,VorlaenderA1932,NioriSWFT1996,AcharyaPK2004,MadsenDNS2004} 
% that attracted much attention during in recent years, belong to this class of our classification.  
% Their phase diagram is indeed so diverse that there are still phases that are not yet understood satisfactorily.
Patchy particles may also belong to the class of colloidal particles without any kind of symmetry. 
This is the case, if the patches are arranged or sized in an irregular way. 
Irregular patchy particles that are made by coating of spherical particles are always convex. 
Colloidal molecules of arbitrary shape and size belong on the other hand to the completely asymmetric colloidal particles 
that are not convex \cite{ManoharanEP2003,QuillietZRvBI2008,SolomonZOSDSBGM2010,KraftVvKvBIK2009}.

\section{\label{sec:derivation}Derivation of the DDFT equation}
In this derivation, we consider a set of $N$ asymmetric rigid particles in a solvent with dynamic (shear) viscosity $\eta$ and neglect possible additional 
(for example vibrational) degrees of freedom.
We choose the center-of-mass positions $\vec{r}_{i}=(x_{1,i},x_{2,i},x_{3,i})$ and the Eulerian angles $\vec{\varpi}_{i}=(\phi_{i},\theta_{i},\chi_{i})$ 
with $i=1,\dotsc,N$ to describe their positions and orientations completely. 
Alternatively, the orientation of the particles could also be described by means of two perpendicular axes \cite{CaldereraFW2004}, but for our purposes, 
the use of Eulerian angles is more appropriate, since they do not involve additional geometric constraints and lead to simpler equations with 
a more compact notation. 
The angular velocities $\vec{\omega}_{i}$ that describe the instantaneous rotational motion of the particles can be expressed in terms of the 
Eulerian angles and their temporal derivatives \cite{Schutte1976}.
For convenience, we use the convention of Gray and Gubbins \cite{GrayG1984}, which is equivalent to the second convention of Schutte \cite{Schutte1976},
for the Eulerian angles, since with this convention, the first two Eulerian angles $\phi$ and $\theta$ are identical to the usual azimuthal and 
polar angles of the spherical coordinate system, respectively.
The whole set of particles is then characterized by the positional and orientational "multivectors" $\vec{r}^{N}=(\vec{r}_{1},\dotsc,\vec{r}_{N})$
and $\vec{\varpi}^{N}=(\vec{\varpi}_{1},\dotsc,\vec{\varpi}_{N})$, respectively. 
For completeness, we also introduce the abbreviation $\vec{\omega}^{N}=(\vec{\omega}_{1},\dotsc,\vec{\omega}_{N})$, here.
The particles are exposed to the (time-dependent) total potential 
\begin{equation}
U(\vec{r}^{N}\!,\vec{\varpi}^{N}\!,t)=U_{\mathrm{ext}}(\vec{r}^{N}\!,\vec{\varpi}^{N}\!,t)+U_{\mathrm{int}}(\vec{r}^{N}\!,\vec{\varpi}^{N})\;,
\end{equation}
which consists of the external potential 
\begin{equation}
U_{\mathrm{ext}}(\vec{r}^{N}\!,\vec{\varpi}^{N}\!,t)=\sum^{N}_{i=1}U_{1}(\vec{r}_{i},\vec{\varpi}_{i},t)
\end{equation}
and the total particle interaction potential
\begin{equation}
U_{\mathrm{int}}(\vec{r}^{N}\!,\vec{\varpi}^{N})
=\sum^{N}_{\begin{subarray}{c}i,j=1\\i<j\end{subarray}}U_{2}(\vec{r}_{i},\vec{r}_{j},\vec{\varpi}_{i},\vec{\varpi}_{j})\;.
\end{equation}
For both the one-particle interaction potentials $U_{1}(\vec{r}_{i},\vec{\varpi}_{i},t)$ and the two-particle interaction potentials 
$U_{2}(\vec{r}_{i},\vec{r}_{j},\vec{\varpi}_{i},\vec{\varpi}_{j})$, we assume pairwise additivity. Moreover, we neglect many-particle interaction potentials of higher order than pair interaction 
potentials. We further introduce the $N$-particle probability distribution function $P(\vec{r}^{N}\!,\vec{\varpi}^{N}\!,t)$ for the probability density
to find the $N$ particles at time $t$ with the orientations $\vec{\varpi}^{N}$ at the positions $\vec{r}^{N}$.
Successive integration of this function with respect to its positional and orientational degrees of freedom leads to the 
$n$-particle density \cite{ArcherE2004}
\begin{equation}
\begin{split}
\rho^{(n)}(\vec{r}^{n}\!,\vec{\varpi}^{n}\!,t)&=\frac{N!}{(N-n)!}\int_{\mathcal{V}}\!\!\dif V_{n+1}\dotsi\!\!\int_{\mathcal{V}}\!\!\dif V_{N} \\
&\times\int_{\mathcal{S}}\!\!\dif\Omega_{n+1}\dotsi\!\!\int_{\mathcal{S}}\!\!\dif\Omega_{N}P(\vec{r}^{N}\!,\vec{\varpi}^{N}\!,t)\;,
\end{split}
\end{equation}
where $\mathcal{V}=\R^{3}$ and $\mathcal{S}=[0,2\pi)\times[0,\pi)\times[0,2\pi)$ are the domains for spatial and orientational 
integration, respectively, $\dif V=\dif x_{1}\dif x_{2}\dif x_{3}$ and $\dif\Omega=\dif\phi\dif\theta\sin(\theta)\dif\chi$ are the corresponding
differentials, and 
\begin{equation}
\begin{split}
\int_{\mathcal{V}}\!\!\dif V&=\int^{\infty}_{0}\!\!\!\!\!\!\dif x_{1}\int^{\infty}_{0}\!\!\!\!\!\!\dif x_{2}\int^{\infty}_{0}\!\!\!\!\!\!\dif x_{3}\;,\\
\int_{\mathcal{S}}\!\!\dif\Omega&=\int^{2\pi}_{0}\!\!\!\!\!\!\!\dif\phi \int^{\pi}_{0}\!\!\!\!\dif\theta\sin(\theta)\int^{2\pi}_{0}\!\!\!\!\!\!\!\dif\chi
\end{split}
\end{equation}
are common abbreviations.

\subsection{Smoluchowski equation}
We start with the derivation of the Smoluchowski equation for the overdamped Brownian dynamics of $N$ self-propelled biaxial particles.
In analogy to the uniaxial passive case (see reference \cite{Dhont1996}), we define the translational gradient operator 
$\Nabla_{\vec{r}}=(\partial_{x_{1}},\partial_{x_{2}},\partial_{x_{3}})$
and the rotational gradient operator $\Nabla_{\vec{\varpi}}=\ii\hat{\mathrm{L}}$, where $\ii$ is the imaginary unit and 
$\hat{\mathrm{L}}=(\mathrm{L}_{x_{1}},\mathrm{L}_{x_{2}},\mathrm{L}_{x_{3}})$ is the angular momentum operator, which can be expressed in terms of
the Eulerian angles by \cite{Schutte1976}
\begin{equation}
\begin{split}
\ii\:\! \mathrm{L}_{x_{1}}&=-\cos(\phi)\cot(\theta)\pdif{}{\phi}-\sin(\phi)\pdif{}{\theta}\\
&\quad\,+\cos(\phi)\csc(\theta)\pdif{}{\chi}\;,\\
\ii\:\! \mathrm{L}_{x_{2}}&=-\sin(\phi)\cot(\theta)\pdif{}{\phi}+\cos(\phi)\pdif{}{\theta}\\
&\quad\,+\sin(\phi)\csc(\theta)\pdif{}{\chi}\;,\\
\ii\:\! \mathrm{L}_{x_{3}}&=\pdif{}{\phi}\;.
\end{split}
\end{equation}
We further define the vectors $\vec{\mathfrak{x}}^{N}=(\vec{r}^{N}\!,\vec{\varpi}^{N})$ and $\vec{\mathfrak{v}}^{N}=(\dot{\vec{r}}^{N}\!,\vec{\omega}^{N})$ 
with $\dot{\vec{r}}^{N}=\dif\vec{r}^{N}\!/\dif t$ and the operators $\Nabla_{\vec{r}^{N}}=(\Nabla_{\vec{r}_{1}},\dotsc,\Nabla_{\vec{r}_{N}})$,
$\Nabla_{\vec{\varpi}^{N}}=(\Nabla_{\vec{\varpi}_{1}},\dotsc,\Nabla_{\vec{\varpi}_{N}})$, and
$\Nabla_{\vec{\mathfrak{x}}^{N}}=(\Nabla_{\vec{r}^{N}},\Nabla_{\vec{\varpi}^{N}})$ and write down the continuity equation
\begin{equation}
\pdif{}{t}P(\vec{\mathfrak{x}}^{N}\!,t)=-\Nabla_{\vec{\mathfrak{x}}^{N}}\!\cdot\big(\vec{\mathfrak{v}}^{N} P(\vec{\mathfrak{x}}^{N}\!,t)\big)\;,
\label{eq:ce}
\end{equation}
which is a trivial generalization of the continuity equation for passive rods that is described by Dhont in Ref.\ \cite{Dhont1996}.
On the Brownian time scale, the total force and torque, acting on an arbitrary particle $i\in\{1,\dotsc,N\}$ are zero.
The total force and torque consist of the force $\vec{F}^{(\mathrm{A})}_{i}(\vec{\mathfrak{x}}^{N}\!,t)$ and 
torque $\vec{T}^{(\mathrm{A})}_{i}(\vec{\mathfrak{x}}^{N}\!,t)$ due to the activity of the self-propelled particle $i$, 
the hydrodynamic force $\vec{F}^{(\mathrm{H})}_{i}(\vec{\mathfrak{x}}^{N})$ 
and torque $\vec{T}^{(\mathrm{H})}_{i}(\vec{\mathfrak{x}}^{N})$, the interaction force $\vec{F}^{(\mathrm{I})}_{i}(\vec{\mathfrak{x}}^{N}\!,t)$ 
and torque $\vec{T}^{(\mathrm{I})}_{i}(\vec{\mathfrak{x}}^{N}\!,t)$ due to the potential $U(\vec{\mathfrak{x}}^{N}\!,t)$, and the Brownian force
$\vec{F}^{(\mathrm{Br})}_{i}(\vec{\mathfrak{x}}^{N}\!,t)$ and torque $\vec{T}^{(\mathrm{Br})}_{i}(\vec{\mathfrak{x}}^{N}\!,t)$.
With the definition $\vec{X}=(\vec{X}_{1},\dotsc,\vec{X}_{N})$ for $\vec{X}\in\{\vec{F}^{(\,\cdot\,)},\vec{T}^{(\,\cdot\,)},\vec{K}^{(\,\cdot\,)}\}$ 
and the abbreviations 
\begin{equation}
\begin{split}
\vec{K}^{(\mathrm{A})}(\vec{\mathfrak{x}}^{N}\!,t)&=(\vec{F}^{(\mathrm{A})}(\vec{\mathfrak{x}}^{N}\!,t),\vec{T}^{(\mathrm{A})}(\vec{\mathfrak{x}}^{N}\!,t))\;,\\
\vec{K}^{(\mathrm{H})}(\vec{\mathfrak{x}}^{N})&=(\vec{F}^{(\mathrm{H})}(\vec{\mathfrak{x}}^{N}),\vec{T}^{(\mathrm{H})}(\vec{\mathfrak{x}}^{N}))\;,\\
\vec{K}^{(\mathrm{I})}(\vec{\mathfrak{x}}^{N}\!,t)&=(\vec{F}^{(\mathrm{I})}(\vec{\mathfrak{x}}^{N}\!,t),\vec{T}^{(\mathrm{I})}(\vec{\mathfrak{x}}^{N}\!,t))\;,\\
\vec{K}^{(\mathrm{Br})}(\vec{\mathfrak{x}}^{N}\!,t)&=(\vec{F}^{(\mathrm{Br})}(\vec{\mathfrak{x}}^{N}\!,t),\vec{T}^{(\mathrm{Br})}(\vec{\mathfrak{x}}^{N}\!,t))\;,
\end{split}
\end{equation}
this force balance for the $N$ colloidal particles can be expressed by
\begin{equation}
\begin{split}
\vec{0}&=\vec{K}^{(\mathrm{A})}(\vec{\mathfrak{x}}^{N}\!,t)+\vec{K}^{(\mathrm{H})}(\vec{\mathfrak{x}}^{N})\\
&\quad+\vec{K}^{(\mathrm{I})}(\vec{\mathfrak{x}}^{N}\!,t)+\vec{K}^{(\mathrm{Br})}(\vec{\mathfrak{x}}^{N}\!,t)\;.
\end{split}
\label{eq:kb}
\end{equation}
The forces and torques resulting from the self-propulsion mechanism of the particles are supposed to be constant with respect to their
orientations in the respective body-fixed coordinate systems, but their strengths may vary slowly with time. 
We denote these forces and torques for a certain particle $i$ in body-fixed Cartesian coordinates by the vector
$\vec{K}^{(\mathrm{A})}_{0,i}(\vec{r}_{i},t)$ for $i=1,\dotsc,N$ and the corresponding vector in space-fixed coordinates by 
\begin{equation}
\vec{K}^{(\mathrm{A})}_{i}(\vec{r}_{i},\vec{\varpi}_{i},t)=\mathcal{R}^{-1}_{0}(\vec{\varpi}_{i})\vec{K}^{(\mathrm{A})}_{0,i}(\vec{r}_{i},t)
\end{equation}
with the diagonal block rotation matrix
\begin{equation}
\mathcal{R}_{0}(\vec{\varpi})=\diag\!\big(\mathrm{R}(\vec{\varpi}),\mathrm{R}(\vec{\varpi})\big)\;,
\end{equation}
where the rotation matrix $\mathrm{R}(\vec{\varpi})$ is defined by
\begin{equation}
\begin{split}
\mathrm{R}(\vec{\varpi})&=\mathrm{R}_{3}(\chi)\,\mathrm{R}_{2}(\theta)\,\mathrm{R}_{3}(\phi)\;,\\
\mathrm{R}^{-1}(\vec{\varpi})&=\mathrm{R}^{\mathrm{T}}(\vec{\varpi})=\mathrm{R}_{3}(-\phi)\,\mathrm{R}_{2}(-\theta)\,\mathrm{R}_{3}(-\chi)
\end{split}
\end{equation}
with the elementary rotation matrices
\begin{equation}
\begin{split}
\mathrm{R}_{2}(\varphi)&=
\begin{pmatrix}
\cos(\varphi)  & 0              & -\sin(\varphi) \\
0              & 1              & 0              \\
\sin(\varphi)  & 0              & \cos(\varphi) 
\end{pmatrix} , \\
\mathrm{R}_{3}(\varphi)&=
\begin{pmatrix}
\cos(\varphi)  & \sin(\varphi)  & 0              \\
-\sin(\varphi) & \cos(\varphi)  & 0              \\
0              & 0              & 1
\end{pmatrix} .
\end{split}
\end{equation}
Note that $\vec{K}^{(\mathrm{A})}_{0,i}$ depends most often only on time $t$, but one could also think of swimming microorganisms in a 
poisoned environment, where $\vec{K}^{(\mathrm{A})}_{0,i}$ also depends on $\vec{r}_{i}$.
To simplify the notation in the following, we collect all the $N$ vectors $\vec{K}^{(\mathrm{A})}_{i}(\vec{r}_{i},\vec{\omega}_{i},t)$
in the vector
\begin{equation}
\vec{K}^{(\mathrm{A})}(\vec{\mathfrak{x}}^{N}\!,t)=\mathcal{R}^{-1}(\vec{\varpi}^{N})\vec{K}^{(\mathrm{A})}_{0}(\vec{r}^{N}\!,t)
\end{equation}
with the $6N\!\times\!6N$-dimensional rotation matrix
\begin{equation}
\mathcal{R}(\vec{\varpi}^{N})=\diag\!\big(\mathcal{R}_{0}(\vec{\varpi}_{1}),\dotsc,\mathcal{R}_{0}(\vec{\varpi}_{N})\big)
\end{equation}
and the $6N$-dimensional vector 
\begin{equation}
\vec{K}^{(\mathrm{A})}_{0}(\vec{r}^{N}\!,t)=\big(\vec{K}^{(\mathrm{A})}_{0,1}(\vec{r}_{1},t),
\dotsc,\vec{K}^{(\mathrm{A})}_{0,N}(\vec{r}_{N},t)\big)\;.
\end{equation}
Next, we focus on the hydrodynamic force and torque. They are given by 
\begin{equation}
\vec{K}^{(\mathrm{H})}(\vec{\mathfrak{x}}^{N})=-\Upsilon(\vec{\mathfrak{x}}^{N})\cdot\vec{\mathfrak{v}}^{N}
\label{eq:hg}
\end{equation}
with the microscopic friction matrix \cite{Dhont1996}
\begin{equation}
\Upsilon(\vec{\mathfrak{x}}^{N})=\begin{pmatrix}\Upsilon^{\mathrm{TT}}(\vec{\mathfrak{x}}^{N})&\Upsilon^{\mathrm{TR}}(\vec{\mathfrak{x}}^{N})\\
\Upsilon^{\mathrm{RT}}(\vec{\mathfrak{x}}^{N})&\Upsilon^{\mathrm{RR}}(\vec{\mathfrak{x}}^{N})\end{pmatrix},
\end{equation}
where $\Upsilon^{\mathrm{TT}}(\vec{\mathfrak{x}}^{N})$, $\Upsilon^{\mathrm{TR}}(\vec{\mathfrak{x}}^{N})$, 
$\Upsilon^{\mathrm{RT}}(\vec{\mathfrak{x}}^{N})$, and $\Upsilon^{\mathrm{RR}}(\vec{\mathfrak{x}}^{N})$ are $3N\!\times\!3N$-dimensional submatrices.
The submatrices $\Upsilon^{\mathrm{TT}}(\vec{\mathfrak{x}}^{N})$ and $\Upsilon^{\mathrm{RR}}(\vec{\mathfrak{x}}^{N})$ correspond to pure translational 
and rotational motion, respectively, while $\Upsilon^{\mathrm{TR}}(\vec{\mathfrak{x}}^{N})$ and $\Upsilon^{\mathrm{RT}}(\vec{\mathfrak{x}}^{N})$
have to be taken into account for particles with a translational-rotational coupling as, for example, screw-like particles. 
For many other particles like those that are orthotropic, however, $\Upsilon^{\mathrm{TR}}(\vec{\mathfrak{x}}^{N})$ and 
$\Upsilon^{\mathrm{RT}}(\vec{\mathfrak{x}}^{N})$ vanish. 
In the following, we neglect hydrodynamic interactions between the colloidal particles.
With this assumption, the microscopic friction submatrices simplify to the block diagonal matrices 
{\allowdisplaybreaks\begin{align}%
\Upsilon^{\mathrm{TT}}(\vec{\varpi}^{N})&=\diag\!\big(\Upsilon^{\mathrm{TT}}_{11}(\vec{\varpi}_{1}),\dotsc,\Upsilon^{\mathrm{TT}}_{NN}(\vec{\varpi}_{N})\big)\;,\\
\Upsilon^{\mathrm{TR}}(\vec{\varpi}^{N})&=\diag\!\big(\Upsilon^{\mathrm{TR}}_{11}(\vec{\varpi}_{1}),\dotsc,\Upsilon^{\mathrm{TR}}_{NN}(\vec{\varpi}_{N})\big)\;,\\
\Upsilon^{\mathrm{RT}}(\vec{\varpi}^{N})&=\diag\!\big(\Upsilon^{\mathrm{RT}}_{11}(\vec{\varpi}_{1}),\dotsc,\Upsilon^{\mathrm{RT}}_{NN}(\vec{\varpi}_{N})\big)\;,\\
\Upsilon^{\mathrm{RR}}(\vec{\varpi}^{N})&=\diag\!\big(\Upsilon^{\mathrm{RR}}_{11}(\vec{\varpi}_{1}),\dotsc,\Upsilon^{\mathrm{RR}}_{NN}(\vec{\varpi}_{N})\big)
\end{align}}%
with the $3\!\times\!3$-dimensional matrices 
{\allowdisplaybreaks\begin{align}%
\Upsilon^{\mathrm{TT}}_{ii}(\vec{\varpi}_{i})&=\eta\,\mathrm{R}^{-1}(\vec{\varpi}_{i})\,\mathrm{K}\,\mathrm{R}(\vec{\varpi}_{i})\;,\\
\Upsilon^{\mathrm{TR}}_{ii}(\vec{\varpi}_{i})&=\eta\,\mathrm{R}^{-1}(\vec{\varpi}_{i})\,\mathrm{C}^{\mathrm{(S)T}}\,\mathrm{R}(\vec{\varpi}_{i})\;,\\
\Upsilon^{\mathrm{RT}}_{ii}(\vec{\varpi}_{i})&=\eta\,\mathrm{R}^{-1}(\vec{\varpi}_{i})\,\mathrm{C}^{\mathrm{(S)}}\,\mathrm{R}(\vec{\varpi}_{i})\;,\\
\Upsilon^{\mathrm{RR}}_{ii}(\vec{\varpi}_{i})&=\eta\,\mathrm{R}^{-1}(\vec{\varpi}_{i})\,\Omega^{\mathrm{(S)}}\,\mathrm{R}(\vec{\varpi}_{i})
\end{align}}%
for $i=1,\dotsc,N$, which are related to the translation tensor $\mathrm{K}$, the coupling tensor $\mathrm{C}^{\mathrm{(S)}}$, 
its transposed $\mathrm{C}^{\mathrm{(S)T}}$, and the rotation tensor $\Omega^{\mathrm{(S)}}$ \cite{HappelB1991} by an orthogonal transformation
with the rotation matrix $\mathrm{R}(\vec{\varpi})$.
The tensors $\mathrm{K}$, $\mathrm{C}^{\mathrm{(S)}}$, and $\Omega^{\mathrm{(S)}}$ are constant and depend on shape and size of the colloidal
particles that are considered, but are independent of the viscosity of the solvent. 
In addition, $\mathrm{C}^{\mathrm{(S)}}$ and $\Omega^{\mathrm{(S)}}$ depend also on the reference point $\mathrm{S}$, for which the center-of-mass 
position of the considered colloidal particle should be chosen. 
In the special case of no hydrodynamic interaction, the inverse of the microscopic friction matrix 
\begin{equation}
\Upsilon^{-1}(\vec{\mathfrak{x}}^{N})=\beta\,\mathcal{D}(\vec{\mathfrak{x}}^{N})
\end{equation}
with the inverse temperature $\beta=1/(k_{\mathrm{B}}T)$, the Boltzmann constant $k_{\mathrm{B}}$, and the microscopic short-time diffusion matrix
\begin{equation}
\mathcal{D}(\vec{\mathfrak{x}}^{N})=\begin{pmatrix}\mathcal{D}^{\mathrm{TT}}(\vec{\mathfrak{x}}^{N})&\mathcal{D}^{\mathrm{TR}}(\vec{\mathfrak{x}}^{N})\\
\mathcal{D}^{\mathrm{RT}}(\vec{\mathfrak{x}}^{N})&\mathcal{D}^{\mathrm{RR}}(\vec{\mathfrak{x}}^{N})\end{pmatrix},
\end{equation}
which we need in the following, has the same structure as the microscopic friction matrix.
We further have the equation 
\begin{equation}
\vec{K}^{(\mathrm{I})}(\vec{\mathfrak{x}}^{N}\!,t)=-\Nabla_{\vec{\mathfrak{x}}^{N}}U(\vec{\mathfrak{x}}^{N}\!,t)
\label{eq:wg}
\end{equation}
for the interaction force and torque. 
Moreover, the Brownian force and torque $\vec{F}^{(\mathrm{Br})}(\vec{\mathfrak{x}}^{N}\!,t)$ and $\vec{T}^{(\mathrm{Br})}(\vec{\mathfrak{x}}^{N}\!,t)$
can be derived from the equilibrium condition 
\begin{equation}
\lim_{t\to\infty}P(\vec{\mathfrak{x}}^{N}\!,t)\propto e^{-\beta U(\vec{\mathfrak{x}}^{N}\!,t)}
\end{equation}
when $\vec{K}^{(\mathrm{A})}(\vec{\mathfrak{x}}^{N}\!,t)$ is neglected and the vector $\vec{\mathfrak{v}}^{N}$ 
in Eq.\ \eqref{eq:ce} is expressed in terms of the vectors $\vec{\mathfrak{x}}^{N}$, $\vec{K}^{(\mathrm{I})}(\vec{\mathfrak{x}}^{N}\!,t)$,
and $\vec{K}^{(\mathrm{Br})}(\vec{\mathfrak{x}}^{N}\!,t)$ with the help of Eq.\ \eqref{eq:kb}.
This results in 
\begin{equation}
\vec{K}^{(\mathrm{Br})}(\vec{\mathfrak{x}}^{N}\!,t)=-\frac{1}{\beta} \Nabla_{\vec{\mathfrak{x}}^{N}} \ln\!\big(P(\vec{\mathfrak{x}}^{N}\!,t)\big)\;.
\label{eq:bg}
\end{equation}
Using Eqs.\ \eqref{eq:kb}, \eqref{eq:hg}, \eqref{eq:wg}, and \eqref{eq:bg}, the Smoluchowski equation 
\begin{equation}
\pdif{}{t}P(\vec{\mathfrak{x}}^{N}\!,t)=\hat{\mathcal{L}}\, P(\vec{\mathfrak{x}}^{N}\!,t)
\label{eq:se}
\end{equation}
with the Smoluchowski operator 
\begin{equation}
\begin{split}
\hat{\mathcal{L}}=\Nabla_{\vec{\mathfrak{x}}^{N}}\!\cdot\Big(&\mathcal{D}(\vec{\mathfrak{x}}^{N})
\cdot\big(\beta\:\!\Nabla_{\vec{\mathfrak{x}}^{N}}U(\vec{\mathfrak{x}}^{N}\!,t)\\
&-\beta\vec{K}^{(\mathrm{A})}(\vec{\mathfrak{x}}^{N}\!,t)
+\Nabla_{\vec{\mathfrak{x}}^{N}}\big)\!\Big)
\end{split}
\label{eq:so}
\end{equation}
follows now directly from the continuity equation \eqref{eq:ce}.

\subsection{DDFT equation}
Next, we proceed in our derivation by applying the integration operator 
$N\!\int_{\mathcal{V}}\!\!\dif V_{2}\dotsi\!\int_{\mathcal{V}}\!\!\dif V_{N}
\int_{\mathcal{S}}\!\!\dif\Omega_{2}\dotsi\!\int_{\mathcal{S}}\!\!\dif\Omega_{N}$
from the left on the Smoluchowski equation \eqref{eq:se} and obtain the expression
\begin{equation}
\begin{split}
\pdif{}{t}\rho(\vec{\mathfrak{x}},t)=\Nabla_{\vec{\mathfrak{x}}}\cdot\Big(\mathcal{D}(\vec{\mathfrak{x}})
\cdot\big(-\beta\vec{K}_{\mathrm{A}}(\vec{\mathfrak{x}},t)\rho(\vec{\mathfrak{x}},t)& \\
+\Nabla_{\vec{\mathfrak{x}}}\rho(\vec{\mathfrak{x}},t)+\beta\rho(\vec{\mathfrak{x}},t)\Nabla_{\vec{\mathfrak{x}}}U_{1}(\vec{\mathfrak{x}},t)
-\beta\bar{K}(\vec{\mathfrak{x}},t)&\big)\!\Big)
\end{split}
\label{eq:rhog}
\end{equation}
with the short-time diffusion tensor 
\footnote{The reason for us to write $\mathcal{D}(\vec{\mathfrak{x}})$ instead of $\mathcal{D}(\vec{\varpi})$ 
in Eqs.\ \eqref{eq:so} and \eqref{eq:rhog} is that one could in principle also describe systems with a space-dependent 
short-time diffusion tensor. This is especially relevant for fluids with a space-dependent viscosity.} 
\begin{equation}
\mathcal{D}(\vec{\varpi})=\begin{pmatrix}\mathcal{D}^{\mathrm{TT}}_{11}(\vec{\varpi})&\mathcal{D}^{\mathrm{TR}}_{11}(\vec{\varpi})\\
\mathcal{D}^{\mathrm{RT}}_{11}(\vec{\varpi})&\mathcal{D}^{\mathrm{RR}}_{11}(\vec{\varpi})\end{pmatrix}
\end{equation}
for the one-particle density $\rho(\vec{\mathfrak{x}},t)\equiv\rho^{(1)}(\vec{\mathfrak{x}},t)$, where we omitted the index $1$ in 
$\vec{r}_{1}$ and $\vec{\varpi}_{1}$ and used the abbreviations 
$\vec{\mathfrak{x}}=(\vec{r},\vec{\varpi})$, $\Nabla_{\vec{\mathfrak{x}}}=(\Nabla_{\vec{r}},\Nabla_{\vec{\varpi}})$,
$\vec{K}_{\mathrm{A}}(\vec{\mathfrak{x}},t)=\vec{K}^{(\mathrm{A})}_{1}(\vec{\mathfrak{x}},t)$, and
$\bar{K}(\vec{\mathfrak{x}},t)=(\bar{F}(\vec{\mathfrak{x}},t),\bar{T}(\vec{\mathfrak{x}},t))$.
When we further introduce the integration operator 
\begin{equation}
\int_{\mathfrak{G}}\!\!\!\dif\mathfrak{V}=\!
\int_{\mathcal{V}}\!\!\!\:\!\dif V\!\int_{\mathcal{S}}\!\!\!\:\!\dif\Omega
\end{equation}
with the total integration domain $\mathfrak{G}=\mathcal{V}\times\mathcal{S}$ and the corresponding differential 
$\dif\mathfrak{V}=\dif V\dif\Omega$, 
the average force $\bar{F}(\vec{\mathfrak{x}},t)$ and torque $\bar{T}(\vec{\mathfrak{x}},t)$ due to the interaction with other particles 
in Eq.\ \eqref{eq:rhog} are given by 
\begin{equation}
\bar{K}(\vec{\mathfrak{x}},t)=-\int_{\mathfrak{G}}\!\!\!\dif\mathfrak{V}'\,
\rho^{(2)}(\vec{\mathfrak{x}},\vec{\mathfrak{x}}'\!,t)\Nabla_{\vec{\mathfrak{x}}}U_{2}(\vec{\mathfrak{x}},\vec{\mathfrak{x}}')\;.
\label{eq:Kqt}
\end{equation}
In equilibrium with $\vec{K}_{\mathrm{A}}(\vec{\mathfrak{x}},t)=\vec{0}$ and $U_{1}=U_{1}(\vec{\mathfrak{x}})$, 
Eq.\ \eqref{eq:rhog} reduces to the first equation of the Bogoliubov-Born-Green-Kirkwood-Yvon hierarchy for 
molecular fluids \cite{GrayG1984}:
\begin{equation}
\beta\bar{K}_{0}(\vec{\mathfrak{x}})=\Nabla_{\vec{\mathfrak{x}}}\rho_{0}(\vec{\mathfrak{x}})
+\beta\rho_{0}(\vec{\mathfrak{x}})\Nabla_{\vec{\mathfrak{x}}}U_{1}(\vec{\mathfrak{x}})\;.
\label{eq:rI}
\end{equation}
Here, a zero in the index of a function denotes the time-independent equilibrium state of this function. 
For example, the function $\rho_{0}(\vec{\mathfrak{x}})$ denotes the equilibrium one-particle density field that corresponds 
to the time-independent prescribed external potential $U_{1}(\vec{\mathfrak{x}})$. 
On the other hand, we have in equilibrium the relation 
\begin{equation}
\begin{split}
\Nabla_{\vec{\mathfrak{x}}}\rho_{0}(\vec{\mathfrak{x}})&+\beta\rho_{0}(\vec{\mathfrak{x}})\Nabla_{\vec{\mathfrak{x}}}U_{1}(\vec{\mathfrak{x}})\\
&\;\;\;=-\beta\rho_{0}(\vec{\mathfrak{x}})\Nabla_{\vec{\mathfrak{x}}}\:\!\Fdif{\mathcal{F}_{\mathrm{exc}}[\rho_{0}(\vec{\mathfrak{x}})]}{\rho_{0}(\vec{\mathfrak{x}})}
\end{split}
\label{eq:rII}
\end{equation}
with the equilibrium Helmholtz excess free-energy functional $\mathcal{F}_{\mathrm{exc}}[\rho_{0}(\vec{\mathfrak{x}})]$.
This relation follows with 
\begin{equation}
\Nabla_{\vec{\mathfrak{x}}}\:\!c^{(1)}_{0}(\vec{\mathfrak{x}})
=\int_{\mathfrak{G}}\!\!\!\dif\mathfrak{V}'\,
c^{(2)}_{0}(\vec{\mathfrak{x}},\vec{\mathfrak{x}}')\Nabla_{\vec{\mathfrak{x}}'}\rho_{0}(\vec{\mathfrak{x}}')\;,
\end{equation}
where $c^{(n)}_{0}(\vec{\mathfrak{x}}_{1},\dotsc,\vec{\mathfrak{x}}_{n})$ is the
$n$-particle direct correlation function in equilibrium, and 
\begin{equation}
c^{(1)}_{0}(\vec{\mathfrak{x}})
=-\beta\Fdif{\mathcal{F}_{\mathrm{exc}}[\rho_{0}(\vec{\mathfrak{x}})]}{\rho_{0}(\vec{\mathfrak{x}})}
\end{equation}
from the more general form 
\begin{equation}
\begin{split}
\Nabla_{\vec{\mathfrak{x}}}\rho_{0}(\vec{\mathfrak{x}})&+\beta\rho_{0}(\vec{\mathfrak{x}})\Nabla_{\vec{\mathfrak{x}}}U_{1}(\vec{\mathfrak{x}})\\
&\;\;\;=\rho_{0}(\vec{\mathfrak{x}})\int_{\mathfrak{G}}\!\!\!\dif\mathfrak{V}'\,
c^{(2)}_{0}(\vec{\mathfrak{x}},\vec{\mathfrak{x}}')\Nabla_{\vec{\mathfrak{x}}'}\rho_{0}(\vec{\mathfrak{x}}')\\
\end{split}
\end{equation}
of Eqs.\ (14) and (16) in reference \cite{Gubbins1980}.
Equations \eqref{eq:rI} and \eqref{eq:rII} lead to the equilibrium relation 
\begin{equation}
\bar{K}_{0}(\vec{\mathfrak{x}})=-\rho_{0}(\vec{\mathfrak{x}})\Nabla_{\vec{\mathfrak{x}}}\:\!
\Fdif{\mathcal{F}_{\mathrm{exc}}[\rho_{0}(\vec{\mathfrak{x}})]}{\rho_{0}(\vec{\mathfrak{x}})}\;,
\end{equation}
which we use instead of Eq.\ \eqref{eq:Kqt} as closure relation for Eq.\ \eqref{eq:rhog} in the time-dependent (non-equilibrium) situation.
A similar \emph{adiabatic approximation} was used in the derivations of the DDFT equations for isotropic \cite{MarconiT1999,ArcherE2004} 
and uniaxial \cite{RexWL2007} colloidal particles. 
The approximation results in the \textbf{generalized DDFT equation} 
\begin{equation}
\begin{split}
\pdif{\rho(\vec{\mathfrak{x}},t)}{t}=\beta\:\!\Nabla_{\vec{\mathfrak{x}}}\cdot
\biggl(\mathcal{D}(\vec{\mathfrak{x}})\cdot\biggl(\rho(\vec{\mathfrak{x}},t)
\biggl(\Nabla_{\vec{\mathfrak{x}}}\:\!
\Fdif{\mathcal{F}[\rho(\vec{\mathfrak{x}},t)]}{\rho(\vec{\mathfrak{x}},t)}&\\
-\vec{K}_{\mathrm{A}}(\vec{\mathfrak{x}},t)\biggr)&\!\biggr)\!\biggr)
\end{split}
\label{eq:BDDFT}
\end{equation}
for anisotropic colloidal particles with the total equilibrium Helmholtz free-energy functional 
\begin{equation}
\mathcal{F}[\rho_{0}(\vec{\mathfrak{x}})]=\mathcal{F}_{\mathrm{id}}[\rho_{0}(\vec{\mathfrak{x}})]
+\mathcal{F}_{\mathrm{exc}}[\rho_{0}(\vec{\mathfrak{x}})]+\mathcal{F}_{\mathrm{ext}}[\rho_{0}(\vec{\mathfrak{x}})]
\end{equation}
that can be decomposed into the ideal rotator gas part \cite{Evans1979} 
\begin{equation}
\beta\mathcal{F}_{\mathrm{id}}[\rho_{0}(\vec{\mathfrak{x}})]=
\int_{\mathfrak{G}}\!\!\!\dif\mathfrak{V}\,
\rho_{0}(\vec{\mathfrak{x}})\big(\ln\!\big(\Lambda^{3}\rho_{0}(\vec{\mathfrak{x}})\big)-1\big)
\end{equation}
with the thermal de Broglie wavelength $\Lambda$, the excess free-energy part 
$\mathcal{F}_{\mathrm{exc}}[\rho_{0}(\vec{\mathfrak{x}})]$, and the contribution \cite{Evans1979} 
\begin{equation}
\mathcal{F}_{\mathrm{ext}}[\rho_{0}(\vec{\mathfrak{x}})]=
\int_{\mathfrak{G}}\!\!\!\dif\mathfrak{V}\,
\rho_{0}(\vec{\mathfrak{x}})U_{1}(\vec{\mathfrak{x}},t)
\end{equation}
due to the external potential $U_{1}(\vec{\mathfrak{x}},t)$.
The DDFT equation \eqref{eq:BDDFT} describes the time evolution of the one-particle density for a system of similar 
anisotropic self-propelled colloidal particles that interact over a pair potential and is the main result of this paper.

\section{\label{sec:applications}Special cases and possible applications}
There is no translational-rotational coupling in the uniaxial case, which means that 
$\mathcal{D}^{\mathrm{TR}}(\vec{\mathfrak{x}}^{N})$ and $\mathcal{D}^{\mathrm{RT}}(\vec{\mathfrak{x}}^{N})$ and therefore also 
$\mathcal{D}^{\mathrm{TR}}_{11}(\vec{\varpi})$ and $\mathcal{D}^{\mathrm{RT}}_{11}(\vec{\varpi})$
vanish in this case.
Furthermore, the one-particle density and the free-energy functional do not depend on the angle $\chi$ for uniaxial particles
and the translational diffusion tensor can then be written as the matrix
$\mathcal{D}^{\mathrm{TT}}(\hat{u})=D_{\parallel}\hat{u}\otimes\hat{u}+D_{\perp}(\Eins-\hat{u}\otimes\hat{u})$, which obviously 
only depends on the two independent short-time diffusion coefficients $D_{\parallel}$ and $D_{\perp}$ for diffusion parallel
and perpendicular to the orientation of the symmetry axis $\hat{u}=(\sin(\theta)\cos(\phi),\sin(\theta)\sin(\phi),\cos(\theta))$
of the uniaxial particle, respectively, where $\Eins$ denotes the $3\!\times\!3$-dimensional unit matrix.
Also the rotational diffusion matrix becomes quite simple for uniaxial particles. 
When we use $\mathcal{D}^{\mathrm{RR}}=D_{\mathrm{R}}\Eins$ with the rotational short-time diffusion coefficient $D_{\mathrm{R}}$
and the considerations above and neglect the self propulsion, we obtain the uniaxial DDFT equation \cite{RexWL2007} from our more 
general DDFT equation \eqref{eq:BDDFT}.
From the uniaxial DDFT equation, one can in turn derive the DDFT equation for two spatial dimensions \cite{WensinkL2008} as well as 
the traditional DDFT equation for colloidal particles with spherical symmetry \cite{MarconiT1999,ArcherE2004} as special cases.

The generalized dynamical density functional theory for passive and active biaxial particles as proposed in Eq.\ \eqref{eq:BDDFT} 
can be numerically solved for a plenty of different problems. 
For passive particles, one can explore for example: 
i) the relaxation dynamics towards equilibrium \cite{RexWL2007}, 
ii) the response of the system to time-dependent external potentials \cite{HaertelBL2010}, 
iii) the growth of a thermodynamically stable phase into an unstable phase \cite{vanTeeffelenBVL2009}. 
Interesting effects for self-propelled particles include among others: 
i) the swarming and clustering behavior of biaxial particles in the bulk and in confinement \cite{PeruaniDB2006,WensinkL2008}, 
ii) the combined impact of self-propulsion and external forcing \cite{WensinkL2008}, 
iii) the effect of space- and time-dependent internal forcing \cite{Rapaport2007}.

\section{\label{sec:conclusions}Conclusions and outlook}
In conclusion, starting from the multi-body Smoluchowski equation, we have derived dynamical density functional
theory for self-propelled Brownian colloidal particles with arbitrary shape. This study was motivated by 
recent progress in synthesizing colloidal particles with (almost) arbitrary shape.
Our results constitute an important framework for further numerical explorations.
This is in particular appealing as since recently an equilibrium density functional is known for arbitrarily shaped hard colloids
\cite{GrafL1999,HansenGoosM2009,HansenGoosM2010} which can serve as an input for the dynamical density functional theory. 
Another possibility to construct a density functional for biaxial particles is a  mean-field approximation for repulsive
segment potentials \cite{RexWL2007}, which works for soft interactions \cite{LikosHLL2002}, 
or a perturbation theory \cite{CurtinA1986,LikosNL1994} for anisotropic attractions around a spherical reference system.
A large number of dynamical problems can then in principle be addressed including the
dynamics \cite{KirchhoffLK1996,Loewen1999,MazzaGVKS2010} and relaxation of nematic-like order in confined systems \cite{RexWL2007,VerhoeffBOvdSL2011}, 
nematic phases driven by external fields \cite{HaertelBL2010}, nucleation kinetics of liquid crystalline phases 
\cite{SchillingF2004,ZhangvD2006,MillerC2009,RavnikZ2009}, and collective behavior of self-propelled particles \cite{PeruaniDB2006,WensinkL2008}. 
The results can be checked against Brownian dynamics computer simulations \cite{Loewen1994b,WhiteCH2001}.

Possible extensions for the future are the inclusions of hydrodynamic interactions 
between the particles which are mediated by the solvent.
Dynamical density functional theory of spherical particles was generalized for hydrodynamic interactions 
\cite{RexL2008,RexL2009,AlmenarR2011}, but this is not yet done for anisotropic particles. Another interesting extension 
would be towards molecular dynamics which is the appropriate dynamics for molecular liquid crystals.
But even for spheres it is much more complicated to formulate a dynamical density functional theory
for molecular dynamics \cite{MarconiTCM2008,Archer2009,EspanolL2009}. 
% Finally, the collective behavior of self-propelled Brownian particles can be described within the framework of 
% dynamical density functional theory. 
% For self-propelled rod-like particles in two spatial dimensions, this was shown in Ref.\ \cite{WensinkL2008}. 
% A generalization towards biaxial particles in three spatial dimensions \cite{WittkowskiL2011} should not pose any major difficulty. 
Finally, the theory can readily be generalized towards binary mixtures \cite{PiniTPR2003}.
%Also the consideration of a given flow field for the solvent would improve DDFT considerably. 

\begin{acknowledgments}
We dedicate this work to Luciano Reatto. We thank Helmut R. Brand, Henricus H. Wensink, Gerhard N{\"a}gele, 
and Joost de Graaf for helpful discussions.
This work has been supported by DFG within SFB TR6 (project D3).
\end{acknowledgments}

\bibliography{References}
\end{document}